\documentclass{aa} 
\usepackage{graphics}
\begin{document} 
 
   \thesaurus{1          % Letters
	      (09.04.1;  % dust, extinction,
               13.09.3;  % Infrared: interstellar: continuum,
               13.09.1)} % Infrared: galaxies.

   \title{Dust Emissivity in the Far-Infrared} 
 
   \subtitle{}
 
   \author{Simone Bianchi \and Jonathan I.\,Davies \and Paul B.\,Alton } 
 
   \offprints{Simone.Bianchi@astro.cf.ac.uk} 
 
   \institute{ Department of Physics \& Astronomy, University of Wales,
   PO Box 913, Cardiff CF2 3YB, U.K. } 
 
   \date{Received  / Accepted } 
 
   \maketitle 
 
   \begin{abstract} 
We have derived the dust emissivity in the Far-Infrared
(FIR) using data available in the literature.
We use two wavelength dependences derived from spectra of 
Galactic FIR emission (Reach et al.~\cite{ReachApJ1995}). 
A value for the emissivity, normalised to the extinction 
efficiency in the V band, has been retrieved from maps of 
Galactic FIR emission, dust temperature and extinction 
(Schlegel et al.~\cite{SchlegelApJ1998}).

Our results are similar to other measurements in the Galaxy
but only marginally consistent with the widely quoted values of
Hildebrand~(\cite{HildebrandQJRAS1983}) derived on one reflection 
nebula. The discrepancy with measurements on other reflection 
nebulae (Casey~\cite{CaseyApJ1991}) is higher and suggests
a different grain composition in these environments with
respect to the diffuse interstellar medium.

We measure dust masses for a sample of six spiral galaxies
with FIR observations and obtain gas-to-dust ratios close
to the Galactic value.

   \end{abstract} 
 
% 
%________________________________________________________________ 
 
\section{Introduction}

Assessing the quantity of dust in spiral galaxies is of primary 
importance in both understanding the intrinsic properties of 
galaxies themselves and interpreting observations of the distant 
universe: large quantities of dust can modify the optical
appearance of galactic structures like spiral arms
(Trewhella \cite{TrewhellaMNRAS1998}); if the distribution of 
dust is extended,
a large fraction of the radiation from the distant universe can be
blocked (Ostriker \& Heisler \cite{OstrikerApJ1984}); star 
formation as determined from UV 
fluxes could be severely underestimated thus altering our knowledge 
of the star formation history of the universe 
(Hughes et al.~\cite{HughesPrep1998}).

Dust mass can be retrieved from extinction or from emission in
the FIR. In the former case information about the star-dust
relative geometry is needed and the method can only be applied to nearby 
edge-on galaxies, where the dust distribution can be
inferred from extinction features (Xilouris et 
al.~\cite{XilourisA&A1997,XilourisA&A1998}).
In the latter case there are no such limitations, and the wealth of
data in the FIR and Sub-mm from instruments like the Sub-mm camera
SCUBA and from the satellites {ISO} and {COBE}, 
can be used to measure dust mass. Unfortunately, the determination of dust 
mass is entangled with that of dust temperature  and they both rely on 
knowledge of the dust emissivity (Hildebrand~\cite{HildebrandQJRAS1983}), 
the form of which is currently highly uncertain.

The emissivity (or emission efficiency, i.e. the ratio between the
emission cross section and the geometric cross section), 
$Q_{\mathrm{em}}(\lambda)$ is usually described by
a function of the form
\begin{equation}
Q_{\mathrm{em}}(\lambda)=Q_{\mathrm{em}}(\lambda_0)
\left(\frac{\lambda_0}{\lambda}\right)^{\beta}
\label{single}
\end{equation}
where $Q_{\mathrm{em}}(\lambda_0)$ is the value of the emissivity at the
reference wavelength $\lambda_0$, and $\beta$ is the wavelength
dependence index.

While a value $\beta=1$ seems to be plausible for $\lambda<100$\,$\mu$m 
(Hildebrand~\cite{HildebrandQJRAS1983}; 
Rowan-Robinson~\cite{RowanRobinsonMNRAS1992}), there is observational 
evidence for a steeper emissivity at longer wavelengths. 
The difference in emissivity is not unexpected, since emission in the 
Mid-Infrared (25-60 $\mu$m) is dominated by transiently heated grains, 
while at $\lambda>100$ grains
emit at thermal equilibrium (Whittet~\cite{WhittetBook1992}).
Sub-mm observations of spiral galaxies 
(Bianchi et al.~\cite{BianchiMNRAS1998}; Alton et al.~\cite{AltonAPJL1998}) 
show that it is not possible to use an emissivity with $\beta=1$
to fit the 450 and 850 $\mu$m emission.
Reach et al.~(\cite{ReachApJ1995}) came to a similar conclusion. 
They used the spectrum of the Galactic plane observed by the 
spectrophotometer FIRAS on board the satellite {COBE}, 
to find that the data are well fitted by an emissivity:
\begin{equation}
Q_{\mathrm{em}}(\lambda)\propto
\frac{
\lambda^{-2}
}{
\left[1+\left(\lambda_1/\lambda\right)^6\right]^{1/6}
},
\label{turning}
\end{equation}
for the range 100 $\mu$m to 1 cm. 
Eq. (\ref{turning}) behaves like (\ref{single}) with $\beta=1$ at
small $\lambda$ ($\lambda\ll\lambda_1$) and $\beta=2$ at large $\lambda$
($\lambda\gg\lambda_1$) (they set $\lambda_1=200$-$\mu$m).

Masi et al.~(\cite{MasiApJ1995}) measure a value 
$\beta=1.54$ by fitting a single 
temperature grey-body spectrum to Galactic plane data in four bands between 
0.5 and 2-mm taken by the balloon born telescope ARGO.
Reach et al.~(\cite{ReachApJ1995}) suggest that a single temperature fit 
may bias 
towards lower values of $\beta$ (see also Wright et al. \cite{WrightApJ1991}); 
over the whole FIRAS spectral range, a two temperature grey-body with 
$\beta=2$ at large $\lambda$ provides a significantly better fit than 
a single temperature spectrum with $\beta\approx 1.5$. 
At long wavelengths theoretical calculations for crystalline substances
constrain $\beta$ to be an even integer number (Wright~\cite{WrightProc1993}).
For amorphous materials $\beta$ depends on the temperature: Agladze et
al.~(\cite{AgladzeApJ1996}) find $1.2<\beta<2$ for amorphous silicate
grains at a temperature of 20~K.

A value for the emissivity at a specific wavelength 
$Q_{\mathrm{em}}(\lambda_0)$ normalised to the extinction efficiency in 
the optical can be determined by carrying out an energy balance in a 
reflection nebula, comparing the energy absorbed from the central star 
with the FIR output from the surrounding dust.  
Alternatively, the extinction measured toward the star can be
directly compared to the optical depth in the FIR 
(Whitcomb et al.~\cite{WhitcombApJ1981}; Hildebrand~\cite{HildebrandQJRAS1983};
Casey~\cite{CaseyApJ1991}).
These methods are complicated by the unknown nebular geometry and by
temperature gradients in the dust; as an example, Casey (\cite{CaseyApJ1991})
found that the extinction method usually retrieves higher values than 
the energy balance.

In this paper we use the extinction method comparing the Galactic
extinction to FIR emission: in this case the same column density 
of dust is responsible both for emission and extinction and a
reliable result can be obtained.

\section{The method}

Schlegel et al.~(\cite{SchlegelApJ1998}; hereafter SFD) have 
presented a new map of Galactic 
extinction. After removing emission from zodiacal light and a cosmic 
infrared background, they have combined the 100 $\mu$m
map of Galactic emission taken by the DIRBE experiment on board the 
{COBE} satellite with the 100 $\mu$m large-area ISSA map from
satellite {IRAS}, to produce
a map of Galactic emission with the quality calibration of DIRBE and the
high resolution of IRAS. The dust temperature has been retrieved using
the DIRBE maps at 100 $\mu$m  and 240 $\mu$m assuming $\beta$=2.
Knowing the temperature, the 100 $\mu$m map has been converted into
a dust column density map and subsequently calibrated to E(B-V) using 
colours and Mg$_2$-index of elliptical galaxies.
We would like to stress that the colour excess has been derived from the 
100 $\mu$m emission without {any} assumption about the value of the 
emissivity at any wavelength. Moreover, the choice of $\beta$ does not 
affect significantly their results: when $\beta$=1.5 is used, the
dust column density map varies only of 1\%, aside from an overall 
multiplicative factor that is taken account of
when calibrating with the colour excess.
We have accessed the electronic distribution of this remarkable dataset 
to retrieve the 9.5$\arcmin$/pixel  maps of the intensity at 100 $\mu$m, 
I(100 $\mu$m), the temperature and the colour excess E(B-V) for the north 
and south Galactic hemispheres. 

When the same dust grains are responsible for emission and extinction, 
the ratio between the extinction coefficient in the V-band and the 
emissivity at 100 $\mu$m is equivalent to the ratio of the optical depths 
\begin{equation}
\frac{Q_{\mathrm{ext}}(V)}{Q_{\mathrm{em}}(\mbox{100 $\mu$m})}=
\frac{\tau(V)}{\tau(\mbox{100 $\mu$m})}.
\label{ratio_tau}
\end{equation}
The above formula is correct if all of the dust grains are identical.
In a mixture of grains of different sizes and materials, the ratio of
emissivities in Eq.~(\ref{ratio_tau}) can still be regarded as a mean 
value characteristic of diffuse galactic dust, if the dust composition 
is assumed to be the same on any line of sight.

The optical depth at 100 $\mu$m, in the optically thin case, is measured using
\begin{equation}
\tau(\mbox{100 $\mu$m})=\frac{I(\mbox{100 $\mu$m})}
{B(\mbox{100 $\mu$m}, T_\mathrm{d})},
\label{taufir}
\end{equation}
where $B(\mbox{100 $\mu$m},T_\mathrm{d})$ is the value of the Planck function at 
$100\mu\mbox{m}$ 
for a dust temperature $T_\mathrm{d}$, both the intensity $I(\mbox{100 $\mu$m})$
and $T_\mathrm{d}$ coming from the maps of SFD.
The optical depth in the V-band can be found from the colour
excess E(B-V) maps,
\begin{equation}
\tau(V)=\frac{A(V)}{1.086}=2.85 E(B-V),
\label{tauopt}
\end{equation}
where we have used a mean galactic value $A(V)/E(B-V)$=3.1
(Whittet~\cite{WhittetBook1992}). Reach et al.~(\cite{ReachApJ1995})
suggest that dust emitting
in the wavelength range 100-300~$\mu$m traces interstellar extinction.
Since the FIR optical depth in eq. (\ref{taufir}) has been measured
using data at 100 and 240 $\mu$m, it is then justified to compare it
with extinction as in eq. (\ref{tauopt}) to find the ratio of the
extinction coefficient and emissivity.
Knowing the optical depths from~(\ref{taufir}) and~(\ref{tauopt}), we can 
compute a map of the ratio as in eq.~(\ref{ratio_tau}); 
we obtain a mean value of
\[
\frac{Q_{\mathrm{ext}}(V)}{Q_{\mathrm{em}}(\mbox{100 $\mu$m})}=760\pm60
\]
for both hemispheres. This value is included, together with other 
multiplicative factors, in the calibration coefficient $p$ as in
eq. (22) in SFD. As pointed out by the referee, an estimate for 
$Q_{\mathrm{ext}}(V)/Q_{\mathrm{em}}$ can be easily derived from 
that equation, if the DIRBE colour corrections factors, slowly 
depending on T, are omitted. Following this way we obtained a value 
of 765.5.  SFD give an error of 8\% for $p$ and this is the 
value quoted here.
Since most ($\approx$ 90\%) of the elliptical galaxies used to calibrate 
colour excess maps have galactic latitude $b>20^\circ$, one may argue that 
the measured value is characteristic only of high latitude dust. 

Reach et al.~(\cite{ReachApJ1995}) 
find that the emissivity (eq.~\ref{turning})
is best determined by fitting the FIRAS spectrum on the Galactic plane. 
They say that
high latitude data have a smaller signal-to-noise ratio and can be fitted
satisfactory with $\beta=2$ (eq.~\ref{single}) although the same 
emissivity as on the plane cannot be excluded. Under the hypothesis that 
the same kind of dust is responsible for the diffuse emission in the whole 
Galaxy, we have corrected SFD temperatures 
using Reach et al. emissivity (eq.~\ref{turning}).
The new temperatures are a few degrees higher than those measured with
$\beta=2$ (as an example we pass from a mean value of 18K in a
20$^\circ$ diameter regions around the north pole to a new estimate 
of 21K\label{newtemp}). It is interesting 
to note that the difference between the two estimates of temperature is of 
the same order as the difference between the temperatures of warm dust at high 
and low Galactic latitude in Reach et al.~(\cite{ReachApJ1995}) and
this may only be a result of the different emissivity used 
to retrieve the temperature.

When the correction is applied 
\[
\frac{Q_{\mathrm{ext}}(V)}{Q_{\mathrm{em}}(\mbox{100 $\mu$m})}=2390\pm190.
\]
The new ratio is about
three times higher, and this is a reflection of the change of
temperature in the black body emission in (\ref{taufir}): for a
higher temperature, a lower emissivity in the FIR is required to 
produce the same emission. Uncertainties in 
$Q_{\mathrm{ext}}(V)/Q_{\mathrm{em}}(\lambda_0)$
are thus greatly affected by assumptions about the emissivity
spectral behaviour.

\section{Comparison with other measurements}
We now compare our emissivity for $\beta=2$ with literature results
derived under the same hypothesis. Since no emissivity has been derived 
to our knowledge assuming eq.~(\ref{turning}), we do not attempt any
comparison with that result. 
All the data are scaled to $\lambda_0=100$ $\mu$m.

Studying the correlation between gas and dust emission from FIRAS and 
DIRBE, Boulanger et al.~(\cite{BoulangerA&A1996}) 
derived an emissivity $\tau/N_H=1.0 \cdot 10^{-25}$ cm$^2$ at 250 $\mu$m
for dust at high galactic latitude; 
assuming the canonical $N_H=5.8 \cdot 10^{21} E(B-V)$ 
cm$^{-2}$ mag$^{-1}$ and $A(V)/E(B-V)$=3.1 (Whittet~\cite{WhittetBook1992}),
this is equivalent to 
$Q_{\mathrm{ext}}(V)/Q_{\mathrm{em}}(\mbox{100 $\mu$m})$=790.
Quite similar values are found in the Draine \& Lee~(\cite{DraineApJ1984}) 
dust model, which has a $\beta=2$ spectral dependence in this
wavelength range. At 125 $\mu$m the optical depth is $\tau/N_H=4.6 
\cdot 10^{-25}$ cm$^2$ which corresponds to 
$Q_{\mathrm{ext}}(V)/Q_{\mathrm{em}}(\mbox{100 $\mu$m})$=680. 
Sodroski et al.~(\cite{SodroskiApJ1997}) 
finds a value for the ratio at 240 $\mu$m, using 
literature data identifying a correlation between B-band extinction and 
100 $\mu$m IRAS surface brightness in high latitude clouds, assuming 
a dust temperature of 18 K. Converted to our notation,
using a standard extinction law, the ratio is 
$Q_{\mathrm{ext}}(V)/Q_{\mathrm{em}}(\mbox{100 $\mu$m})$=990.

The measurement by Whitcomb et al.~(\cite{WhitcombApJ1981}) 
on the reflection nebula NGC 7023 is the most commonly quoted value for
the emissivity (Hildebrand~\cite{HildebrandQJRAS1983}).
Their value derived at 125 $\mu$m for $\beta=2$ is only marginally 
consistent with our result. Following our notation, their result is
equivalent to $Q_{\mathrm{ext}}(V)/Q_{\mathrm{em}}(\mbox{100 $\mu$m})=$ 
220 and 800\footnote{Whitcomb et al.~(\cite{WhitcombApJ1981}) and 
Casey~(\cite{CaseyApJ1991})
originally presented values for $Q_{\mathrm{ext}}(UV)/Q_{\mathrm{em}}(FIR)$: 
we have corrected to $Q_{\mathrm{ext}}(V)/Q_{\mathrm{em}}(FIR)$ using the 
provided $\tau(UV)=2\tau(V)$.}, using the energy balance and the extinction 
method, respectively.
The values obtained by Casey~(\cite{CaseyApJ1991}) on a sample of 
five nebulae using the energy balance method are a factor of 3 smaller 
than ours (corresponding to $Q_{\mathrm{ext}}(V)/Q_{\mathrm{em}}
(\mbox{100 $\mu$m})=$ 80-400).

\begin{figure}
\resizebox{\hsize}{!}{\includegraphics{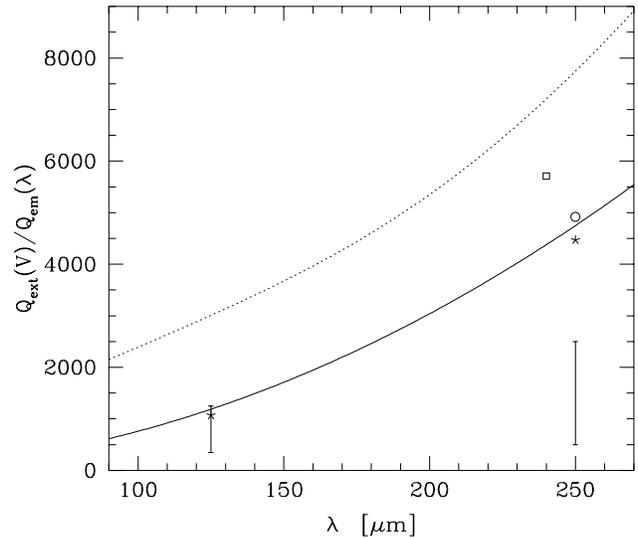}}
\caption{FIR emissivity derived in this work: assuming $\beta$=2 (solid
line) and the Reach et al.~(\cite{ReachApJ1995}) 
spectral dependence (dotted line).
Error bars at 125 $\mu$m give the range of values of
Whitcomb et al.~(\cite{WhitcombApJ1981}).
The error bar at 250 $\mu$m gives the range of values of 
Casey~(\cite{CaseyApJ1991}). 
Data points are from Draine \& Lee~(\cite{DraineApJ1984})
model (stars), Sodroski et al.~(\cite{SodroskiApJ1997})
(square) and Boulanger et al.~(\cite{BoulangerA&A1996}) (circle).}
\label{fig_emissi}
\end{figure}

In Fig.~\ref{fig_emissi} we show the literature data (plotted at the 
wavelength they have been derived in the original papers) together
with our derived emissivity laws.
We have added the value for Draine \& Lee~(\cite{DraineApJ1984})
model at 250 $\mu$m.

\begin{table*}
\begin{center}
\begin{tabular}{lccrcr}
\hline
Galaxy & Gas Mass &  \multicolumn{2}{c}{emissivity~(\protect{\ref{single}})}
&  \multicolumn{2}{c}{emissivity~(\protect{\ref{turning}})}\\
~&10$^{10}$ M$_\odot$& T(K)   & G/D ratio &   T(K)  & G/D ratio \\
\hline
NGC 628     & 1.1  & 16 (17) &  90 (190) & 18 (20) & 100 (200)\\
NGC 660     & 0.91 & 19 (21) & 110 (230) & 23 (26) & 120 (250)\\
NGC 5194    & 0.75 & 18 (20) &  90 (180) & 21 (24) &  90 (190)\\
NGC 5236    & 3.5  & 19 (21) & 240 (500) & 22 (25) & 255 (540)\\ 
NGC 6946    & 3.0  & 17 (19) &  75 (150) & 20 (22) &  80 (160)\\
NGC 7331    & 1.0  & 17 (19) &  70 (145) & 20 (22) &  70 (155)\\
\hline
\end{tabular}
\end{center}
\caption{Sample of galaxies from Alton et al.~(\cite{AltonA&A1998}). 
Gas masses have been derived from Devereux \& Young~(\cite{DevereuxApJ1990})
(Van Driel et al.~(\cite{VanDrielAJ1995}) for NGC 660) and corrected to 
the distances quoted by Alton et al.~(\cite{AltonA&A1998}). 
Dust temperature and gas-to-dust ratio are derived using eq.~(\ref{single}) 
with $\beta$=2 ($Q_{\mathrm{ext}}(V)/Q_{\mathrm{em}}(\mbox{100 $\mu$m})$=760), 
and eq.~(\ref{turning}) 
($Q_{\mathrm{ext}}(V)/Q_{\mathrm{em}}(\mbox{100 $\mu$m})$=2390). 
Values within brackets are derived under
the hypothesis that {ISO} fluxes are overestimated by 30\%.
}
\label{gd_ratio}
\end{table*}

\section{Gas-to-dust ratio of external spiral galaxies}
We now exploit the FIR emissivity derived in this work by
determining dust masses for nearby spiral galaxies.
Following Hildebrand~(\cite{HildebrandQJRAS1983})
dust masses can be measured from FIR emission using
\begin{equation}
M_{\mathrm{dust}}=\frac{F(\lambda) D^2}{B(\lambda,
T_\mathrm{d})}\cdot \frac{4 a \rho}{3 Q_{\mathrm{em}} (\lambda)},
\label{dmass}
\end{equation}
where $F(\lambda)$ is the total flux at the wavelength $\lambda$, $D$ 
the distance of the
object, $B(\lambda,T_\mathrm{d})$ the Planck function, $a$ the grain radius 
(0.1~$\mu$m) and $\rho$ the grain mass density (3~g~cm$^{-3}$). 
The emissivity $Q_{\mathrm{em}}(\lambda)$ is derived from the ratio
$Q_{\mathrm{ext}}(V)/Q_{\mathrm{em}}(\lambda)$ assuming 
$Q_{\mathrm{ext}}(V)$=1.5 (Casey~\cite{CaseyApJ1991};
Whittet~\cite{WhittetBook1992}).

Alton et al.~(\cite{AltonA&A1998}) provide total fluxes at 100 
$\mu$m and 200 $\mu$m from {IRAS} and  {ISO} for a sample of 
spiral galaxies. We have derived dust temperatures and masses using
$Q_{\mathrm{ext}}(V)/Q_{\mathrm{em}}(\mbox{100 $\mu$m})$=760 and 2390,
for $\beta$=2 and Reach et al.~(\cite{ReachApJ1995}) emissivities,
respectively.
Using literature values for gas masses, we have computed the
gas-to-dust ratios. Values of gas masses, temperatures and gas-to-dust
ratios are presented in Table~\ref{gd_ratio}.

The mean value of the gas-to dust ratio for the sample is 100 using 
eq.~(\ref{single}), 110 using eq.~(\ref{turning}).
Mean temperatures go from 18K with the $\beta=2$ emissivity to 
21K when the Reach et al.~(\cite{ReachApJ1995}) behaviour is assumed 
(as for the north galactic pole in Sect.~\ref{newtemp}). 
Alton et al.~(\cite{AltonA&A1998}) pointed out that {ISO} 
200 $\mu$m fluxes could be overestimated by about 30\%; 
correcting for this we obtain a mean gas-to-dust
ratio of 220-240 (for $\beta$=2 and Reach et al.~(\cite{ReachApJ1995})
emissivity, respectively).
As shown above, dust masses obtained with the two methods are quite 
similar. This can be explained substituting eqs.~(\ref{ratio_tau}) and 
(\ref{taufir}) into (\ref{dmass}). For $\lambda=\mbox{100 $\mu$m}$
we can derive
\[
M_{\mathrm{dust}}\sim\frac{B(\mbox{100 $\mu$m},T_\mathrm{d}^\mathrm{G})}
{B(\mbox{100 $\mu$m},T_\mathrm{d})},
\]
where $T_\mathrm{d}^\mathrm{G}$ is the mean temperature of dust in the
Galaxy. From the equation it is clear that the dust mass determination is
insensitive to the emissivity law used, as long as the dust temperature 
in external galaxies and in our own are similar.

Our range of values for the gas-to-dust ratio (100--230) encompasses
the Galactic value of 160 (Sodroski et al.~\cite{SodroskiApJ1994}). 
As a comparison, the mid-value of Whitcomb et al.~(\cite{WhitcombApJ1981}) 
would have given dust-to-gas ratios larger by a factor 1.5.

\section{Conclusion}
We have derived the dust emissivity $Q_{\mathrm{em}}$ in the FIR using 
the wavelength dependence derived from the FIR Galactic spectrum 
(Reach et al.~\cite{ReachApJ1995}). The emissivity has been normalised to the
extinction efficiency in the V band using dust column density maps
calibrated to Galactic extinction (SFD).
$Q_{\mathrm{em}}$ depends strongly on the assumed wavelength dependence.
For a $\beta=2$ emissivity index we obtained
\[
Q_{\mathrm{em}}(\lambda)=\frac{Q_{\mathrm{ext}}(V)}{760}
\left(\frac{\mbox{100 $\mu$m}}{\lambda}\right)^2.
\]
This result is consistent with other values derived from FIR Galactic
emission (Boulanger et al.~\cite{BoulangerA&A1996}; Sodroski et
al.~\cite{SodroskiApJ1997}) and with the Draine \& Lee~(\cite{DraineApJ1984})
dust model.
The widely quoted emissivities of Whitcomb et al.~(\cite{WhitcombApJ1981}; 
Hildebrand~\cite{HildebrandQJRAS1983})
derived from the reflection nebula NGC 7023 are only marginally consistent
with our values, while the emissivity measured by Casey~(\cite{CaseyApJ1991})
on a sample of five nebulae are smaller by a factor of 3.
This may suggest a different grain composition for dust in the diffuse
inter-stellar medium compared to reflection nebulae.

When the wavelength dependence derived by Reach et al.~(\cite{ReachApJ1995})
on the Galactic plane is used, we obtain
\[
Q_{\mathrm{em}}(\lambda)=\frac{Q_{\mathrm{ext}}(V)}{2390}
\left(\frac{\mbox{100 $\mu$m}}{\lambda}\right)^2
\frac{2.005}{
\left[1+\left(\mbox{200 $\mu$m}/\lambda\right)^6\right]^{1/6}
}.
\]
We have used the derived emissivities to measure dust masses from 
100 $\mu$m and 200 $\mu$m fluxes of a sample of six spiral 
galaxies (Alton et al.~\cite{AltonA&A1998}). 
We have retrieved similar dust masses with both the spectral 
dependences. The gas-to dust ratios of our sample (100-230) are 
close to the Galactic value of 160 (Sodroski et al.~\cite{SodroskiApJ1994}).

\begin{acknowledgements} 
We thank M. Edmunds and S. Eales for stimulating discussions,
and an anonymous referee for useful comments that have 
improved the paper.
\end{acknowledgements}

\end{document}